\documentclass[a4paper,oneside,12pt]{article}

\usepackage{setspace}
\usepackage{graphicx}
\usepackage{amsmath}
\usepackage{amssymb}

\begin{document}

\title{Tunneling of interacting fermions in 1D systems}

\author{A. Gendiar, M. Mo\v{s}ko, P. Vagner, and R. N\'{e}meth\medskip\\
{\it Institute of Electrical Engineering, Slovak Academy of Sciences}\\
{\it D\'{u}bravsk\'{a} cesta 9, SK-841 04 Bratislava, Slovakia}}

\date{}
\renewcommand{\abstractname}{}

\maketitle \vspace{-1.cm}
\begin{abstract}
Using the self-consistent Hartree-Fock approximation for spinless
electrons at zero temperature, we study tunneling of the
interacting electron gas through a single $\delta$ barrier in a
finite one-dimensional (1D) wire connected to contacts. Our
results exhibit features known from correlated many-body models.
In particular, the conductance decays with the wire length as
$\propto L^{-2\alpha}$, where the power $\alpha$ is universal. We
also show that a similar result for a wire conductance can be
extracted from the persistent current ($I$) through the $\delta$
barrier in a 1D ring, where it is known that $I \propto
L^{-1-\alpha}$.
\end{abstract}

PACS numbers: 73.23.-b, 73.61.Ey


\medskip
\noindent

Electron gas in a quantum wire is a realistic one-dimensional (1D)
system. If a clean wire is biased by two macroscopic contacts with
negligible electron back-scattering, the wire conductance is
quantized as an integer multiple of $e^2/h$. This effect can be
explained in the model of noninteracting electrons
\cite{Datta-95}.

If a localized scatterer is introduced into the wire, quantization
of conductance breaks down due to the electron backscattering from
the scatterer. For non-interacting electrons, the conductance is
given by Landauer formula as the electron transmission probability
at the Fermi level \cite{Datta-95}. However, the electron-electron
(e-e) interaction alters the properties of the system
qualitatively. In the Luttinger liquid model \cite{Kane-92}, the
conductance of an infinite wire containing a single scatterer
varies with temperature as $\propto T^{-2\alpha}$ for
$T\rightarrow0$, where the power $\alpha$ depends only on the e-e
interaction. For repulsive interaction $\alpha$ is positive and
reflection at zero temperature is perfect, no matter how strong or
weak the scatterer is.

Matveev et al. \cite{Matveev-93} studied the Landauer conductance
of the interacting 1D electron gas through a $\delta$ barrier in a
wire with contacts. They replaced the many-body wave function by
the Slater determinant of single-electron wave functions and
analyzed the effect of the Hartree-Fock potential on the tunneling
transmission. Assuming a weak e-e interaction of finite range,
they derived the transmission using the renormalization group
(RG). They confirmed the universal power law $T^{-2\alpha}$, this
approach is believed to go beyond the Hartree-Fock approximation.

Here we consider the non-Luttinger liquid model of Matveev et
al.~\cite{Matveev-93}, but instead of the RG approach we apply the
self-consistent Hartree-Fock solution. We evaluate the Landauer
conductance. We find a good agreement with the theory of Matveev
et al. In particular, we simulate asymptotic dependence of the
conductance on the wire length ($L$) for strong $\delta$ barriers
and we reproduce the universal power law $\propto L^{-2\alpha}$.
We also show that essentially the same wire conductance can be
extracted from the persistent current ($I$) in a 1D ring, where $I
\propto L^{-\alpha-1}$.

We consider a 1D gas of interacting spinless electrons in a wire
of length $L$. The wire is positioned along the $x$ axis between
$x=-L/2$ and $x=L/2$, both wire ends are connected to contacts.
The single-electron wave functions $\psi_k(x)$, where $k$ is the
electron wave vector, are described by the Hartree-Fock equation
\begin{equation} \label{Schr}
\left[
  -\frac{\hbar^2}{2m}\ \frac{d^2}{dx^2} + \gamma\delta(x) + U_H(x) + U_F(k,x)
\right] \psi_k(x)
\\ =\varepsilon_k \, \psi_k(x),
\end{equation}
where $\gamma\delta(x)$ mimics the localized scatterer positioned
in the center of the wire,
\begin{equation} \label{U_H}
U_H(x)= \\
\int \limits _{-L/2} ^{L/2} dx' \ V(x-x') \int \limits_{-k_F}
^{k_F} \frac{dk'}{2\pi} \left[\left| \psi_{k'}(x') \right|^2 -
\left| \psi^0_{k'}(x') \right|^2 \right]
\end{equation}
is the Hartree potential induced by the $\delta$ barrier,
\begin{equation} \label{U_F1}
U_F(k,x) = -\frac{1}{\psi_k(x)} \int \limits _{-L/2} ^{L/2} \!\!
dx' \ V(x-x') \,
\\ \int \limits _{-k_F} ^{k_F} \!\! \frac{dk'}{2\pi} \, \,
\psi_k(x') \psi^{*}_{k'}(x') \, \psi_{k'}(x)
\end{equation}
is the Fock nonlocal exchange term, and $V(x-x')$ is the e-e
interaction.

The Landauer conductance is defined for the wire
connected to large contacts via adiabatically tapered
non-reflecting connectors \cite{Datta-95}. First assume a clean
non-interacting wire, i.e., $\gamma = 0$ and $V(x-x')=0$. As there
is no backscattering at the wire ends, the solution of the
equation~\eqref{Schr} is the free wave $e^{ikx}$ with eigenenergy
$\hbar^2k^2/2m$. The states $e^{ikx}$ and $e^{-ikx}$ with $k>0$
describe the ballistic electrons originating from the contact at
$x=-L/2$ and $x=L/2$, respectively. As they are mutually
incoherent, $e^{ikx}$ and $e^{-ikx}$ are the only independent
solutions.

Second, let us keep $\gamma = 0$ but let us turn on the e-e
interaction $V(x-x')$. Assuming that there is no backscattering at
the wire ends, the solution of equation \eqref{Schr} is still the
free wave, $\psi^0_{k}(x) = e^{ikx}$, but with the eigenenergy
\begin{equation} \label{e_k}
\varepsilon_k = \hbar^2k^2/2m + U_F^0 (k),
\end{equation}
where $U_F^0 (k)\equiv U_F[\psi_{k}(x)= \psi^0_{k}(x)]$ is the
Fock shift. Note that this solution is valid if we implicitly
assume that the Fock interaction is present also in the contacts.
Indeed, if the energy \eqref {e_k} holds inside the wire and we
turn off the Fock shift to zero outside the wire, we obtain at
each wire end the potential drop $U_F^0 (k)$. This would cause
backscattering at both wire ends and the solutions $e^{ikx}$ and
$e^{-ikx}$ would be no longer valid, in contrast with the
ballistic conductance of clean wires \cite{Datta-95}.

Finally, consider the barrier $\gamma\delta(x)$ and interaction
$V(x-x')$. The $\delta$ barrier induces Friedel oscillations of
the Hartree-Fock potential. The electrons are thus scattered by
the $\delta$ barrier and by the oscillating potential relief.
Since the scattering is elastic, the eigenenergy \eqref {e_k}
remains unchanged and the wave function $\psi_k(x)$ can be found
by solving equation \eqref{Schr} as a tunneling problem with
boundary conditions
\begin{equation} \label{BCondgt}
\psi_k(x=-L/2)=e^{ikx}+r_k e^{-ikx}, \ \quad \quad \quad
\psi_k(x=L/2)=t_k e^{ikx},
\end{equation}
\begin{equation} \label{BCondls}
\psi_{-k}(x = - L/2)=t'_k e^{-ikx}, \ \quad \psi_{-k}(x =
L/2)=e^{-ikx}+r'_k e^{ikx} ,
\end{equation}
where $k >0$, $r_k$ is the reflection amplitude, and $t_k$ is the
transmission amplitude. Once we know the transmission, we know the
Landauer conductance $(e^2/h) \left| t_{k_F} \right|^2$.

In reality the Friedel oscillations penetrate through the wire
ends into the contacts, where they decay fast due to the enhanced
dimensionality and decoherence. To mimic this decay within our 1D
model, we sharply turn off the oscillations to zero at both wire
ends and keep $U_H=0$ and $U_F=U_F^0 (k)$ outside the wire. This
constant 1D potential emulates the non-reflecting connectors and
justifies the above boundary conditions. Essentially the same 1D
model was a starting point of the RG study by Matveev et al.
\cite{Matveev-93}.

Unlike to Matveev et al., we solve the equation \eqref{Schr} by
means of the self-consistent iterative procedure. To save the
computational time and memory, we follow Ref. \cite{Cohen-97} and
simplify the equation \eqref{U_F1} as
\begin{equation} \label{U_F2}
U_F(x) \simeq -\int \limits _{-L/2} ^{L/2} dx' \ V(x-x') \, \int
\limits _{-k_F} ^{k_F} \frac{dk'}{2\pi} \, \text{Re}
[\psi^{*}_{k'}(x') \, \psi_{k'}(x)]
\end{equation}
by noticing that $\int _{-k_F} ^{k_F} dk' \psi^{*}_{k'}(x') \,
\psi_{k'}(x) \simeq 2\pi \delta(x-x')$. Unlike the exact form
\eqref{U_F1}, the Fock potential \eqref{U_F2} is local and
independent on $k$. This saves time and allows us to simulate long
wires. We present numerical results for the GaAs wire with
electron density $n=5 \times 10^7$ m$^{-1}$, effective mass
$m=0.067$~$m_0$, and e-e interaction
\begin{equation} \label{VeeExp}
V(x - x') = V_0 \,  e^{- \left| x - x' \right|/d}.
\end{equation}
We adopt the finite-ranged interaction (\ref{VeeExp}) because of
comparison with the RG theory of Ref.~\cite{Matveev-93} which
also assumes the e-e interaction of finite range. Physical meaning
of the finite range is screening.

According to Ref. \cite{Matveev-93} the bare $\delta$ barrier,
described by the transmission and reflection amplitudes
$\tilde{t}_k$ and $\tilde{r}_k$, is renormalized by the Friedel
oscillations. The renormalized transmission probability at the
Fermi level reads

\begin{equation} \label{t-Glazman-Fermi}
\left| {t}_{k_F} \right|^2 = \frac{\left| \tilde{t}_{k_F}
\right|^2 \, (d/L)^{2\alpha}}
     {{ \left| \tilde{r}_{k_F} \right|^2 +
      \left| \tilde{t}_{k_F} \right|^2 \,
      (d/L)^{2\alpha} }
}\simeq \frac{\left| \tilde{t}_{k_F} \right|^2}{|
\tilde{r}_{k_{F}}|^2}\left(d/L\right)^{2\alpha},
\end{equation}
where $d$ is the range of the e-e interaction $V(x-x')$ and the
right hand side of (\ref{t-Glazman-Fermi}) holds for small
$\tilde{t}_{k_F}$ and/or large $L$. For weak e-e interaction
($\alpha \ll 1$) $\alpha$ reads

\begin{equation} \label{alpha-Glazman}
\alpha = \frac{V(0)-V(2k_F)} {2\pi \hbar v_F},
\end{equation}
where $V(q)$ is the Fourier transform of the e-e interaction
$V(x-x')$. We evaluate $\alpha$ for our e-e interaction
\eqref{VeeExp}, for which $V(q) = 2V_0 d/(1+q^2d^2)$.

The bare amplitudes are $\tilde{t}_k = k/(k+i\zeta)$ and
$\tilde{r}_k = -i\zeta/(k+i\zeta)$, where $\zeta = \gamma
m/\hbar^2$. Since $k_F$ and $m$ are fixed, in the following we
parametrize the bare $\delta$ barrier by its transmission
coefficient $\left| \tilde{t}_{k_F} \right|^2$.

\begin{figure} [tb]
\begin{center}
\begin{minipage}[c]{6cm}
\begin{center}
\includegraphics[clip,width=6cm]{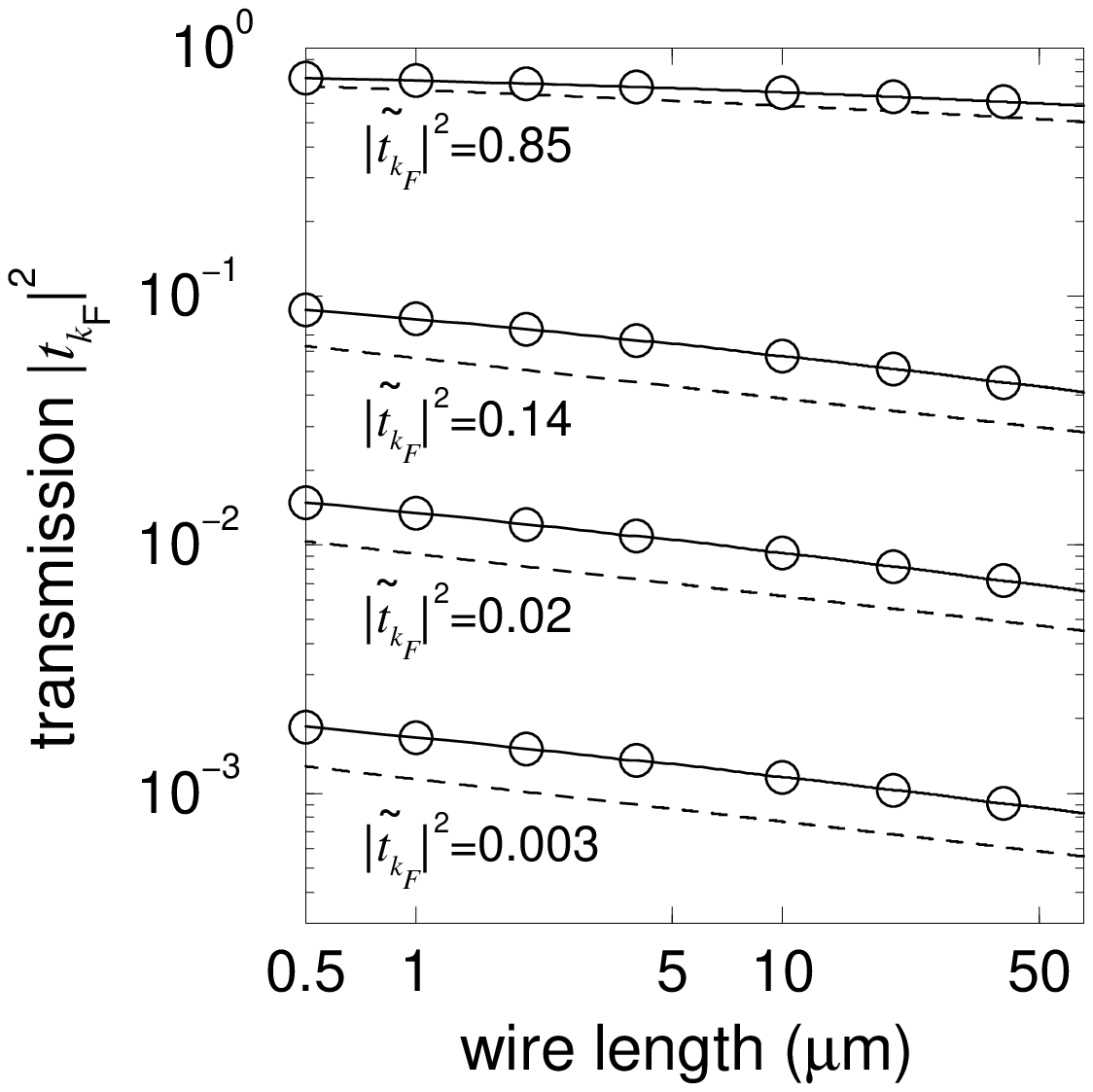}
\end{center}
\end{minipage}
\hspace{0.01cm}
\begin{minipage}[c]{6cm}
\begin{center}
\includegraphics[clip,width=5cm]{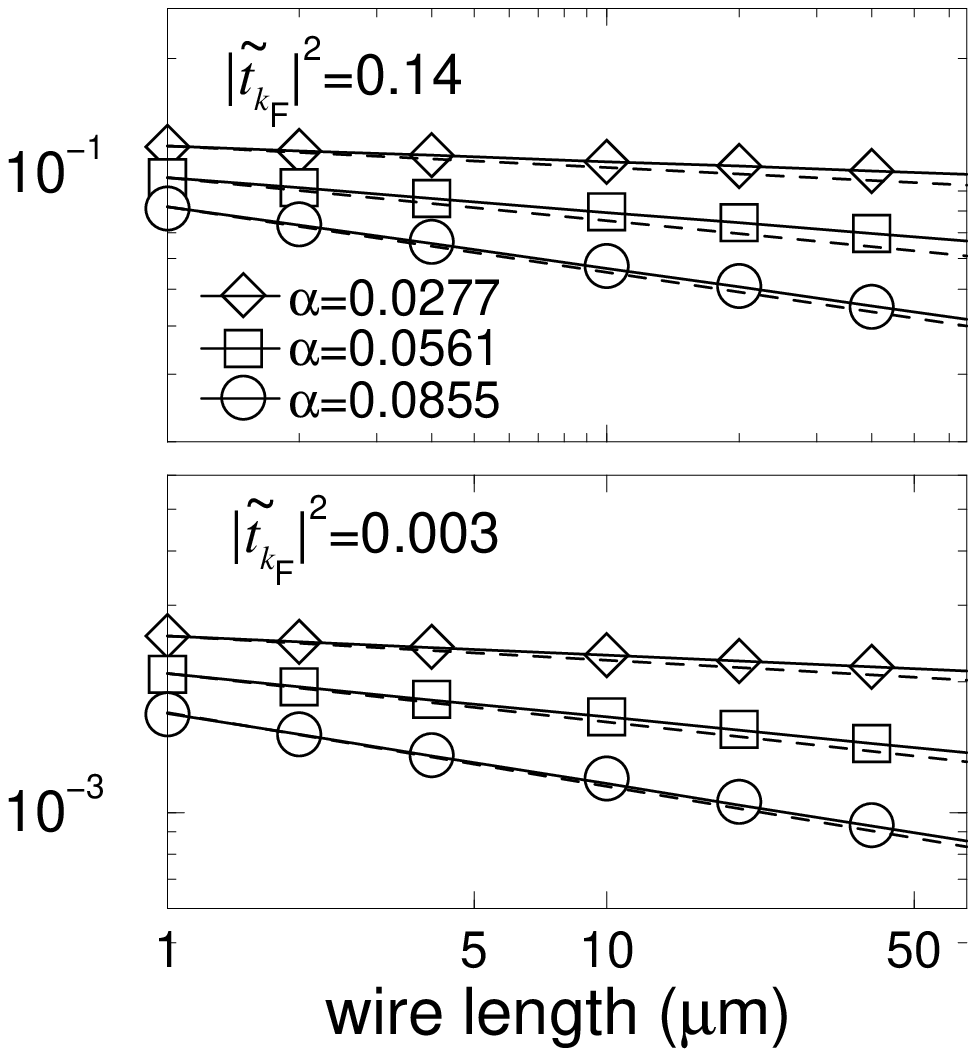}
\end{center}
\end{minipage}
\end{center}
\caption{{\small {Transmission probability at the Fermi level versus
wire length for various $\delta$ barriers $\left| \tilde{t}_{k_F}
\right|^2$ and various e-e interaction strengths $\alpha$. The
dashed curves show the data due to the RG formula
\eqref{t-Glazman-Fermi}; these data follow the asymptotic power
law $\left| t_{k_F} \right|^2 \propto L^{-2\alpha}$ as the
$\delta$ barrier becomes strong. The circles connected by the full
lines are our self-consistent Hartree-Fock data; they approach the
same asymptotic power law. The powers  $\alpha = 0.0277$,
$0.0561$, and $0.0855$ correspond to $V_0 = 11$~meV, $22.3$~meV,
and $34$~meV, respectively, with $d$ fixed to $3$ nm. If we adjust
the same $\alpha$ by another choice of $V_0$ and $d$, our results
remain unchanged.}}} \label{Fig:1}
\end{figure}

Figure \ref{Fig:1} shows the transmission probability $\left|
{t}_{k_{F}} \right|^2$ versus the wire length $L$ for various
$\delta$ barriers and various e-e interaction strengths. The RG
formula~\eqref{t-Glazman-Fermi} is presented by the dashed lines. For
strong $\delta$ barriers the dashed lines follow the asymptotic
power law $\left| t_{k_F} \right|^2 \propto L^{-2\alpha}$, in the
log scale manifested by linear decay with slope $-2\alpha$. Our
Hartree-Fock curves (open symbols connected by full lines) show
slightly higher transmission but clearly follow the same trend. In
particular, for small enough $\left| \tilde{t}_{k_F} \right|^2$
and for not too small $L$ all Hartree-Fock curves decay with the
same slope as the RG curves, independently on the strength of the
$\delta$ barrier.

It is instructive to show the Hartree-Fock potential
$U_H(x)+U_F(x)$ in the form
\begin{equation}\label{HF-normalized}
  U_{HF}(y)= (d/L)^{\alpha}[U_H(y)+U_F(y)-U_F^0]/\Delta_L \,, \quad
y=x/L \,, \quad \Delta_L= \pi \hbar v_F/L
  ,
\end{equation}
where we subtract the constant Fock shift $U_F^0$ in order to show
exclusively the Hartree-Fock potential induced by the $\delta$
barrier.

\begin{figure}[tb]
\begin{center}
\includegraphics[clip,width=10.5cm]{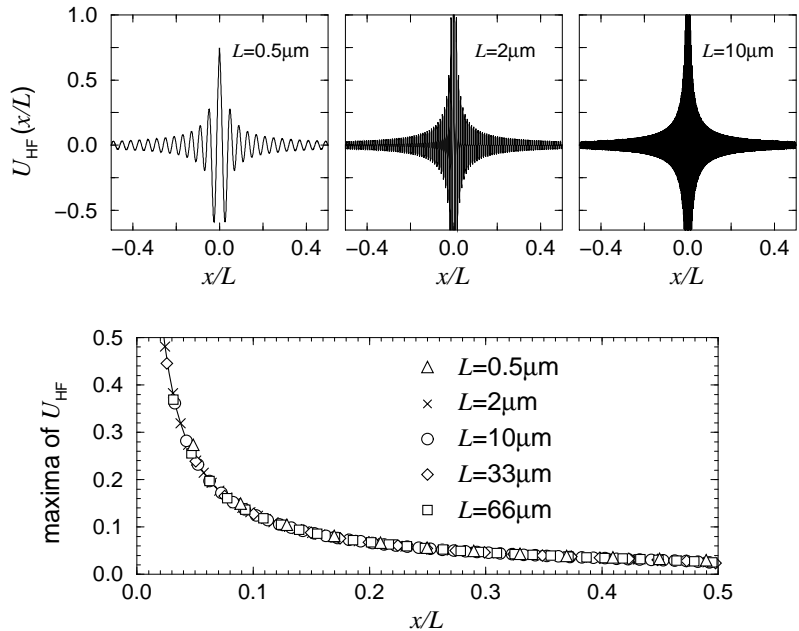}
\end{center}
\vspace{-0.5cm}
\caption{{\small {Self-consistent Hartree-Fock potential
$U_{HF}(x/L)$ along a 1D wire, induced by the (not shown) $\delta$
barrier at $x=0$. The $\delta$ barrier is adjusted to have the
transmission $\left| \tilde{t}_{k_F} \right|^2=0.003$ at the Fermi
level ($14$ meV). The wire length, $L$, is varied as a parameter.
The bottom panel shows just the positive maxima of the Friedel
oscillations.}}} \label{Fig:2}
\end{figure}

In Fig.~\ref{Fig:2} we show the typical self-consistent
$U_{HF}(y)$ in the wire with a strong scatterer at $y=0$. The
potential exhibits Friedel oscillations with period
$\lambda_F/2L$. The bare scatterer is thus ``dressed" by an extra
scatterer due to the Friedel oscillations. This is why we see the
conductance to decay with $L$. It is more difficult to understand
why we see just $\left| t_{k_F} \right|^2 \propto
(d/L)^{2\alpha}$. As $L$ increases, the Friedel oscillations in
Fig.~\ref{Fig:2} are too dense to be distinguishable, but we can
observe asymptotic decay of the oscillation amplitude. Notice that
the ``envelope" of the oscillation amplitude is the same for all
$L$. Indeed, as shown in the bottom panel, the ``envelope" scales
for all $L$ to a single curve. Notice also (c.f.
eq.~\eqref{HF-normalized}), that $U_{HF}(y)$ involves the scaling
factor $(d/L)^{\alpha}$. This might be the reason why $\left|
t_{k_F} \right|^2 \propto (d/L)^{2\alpha}$, but we have so far not
found a clear interpretation.

By tying the wire ends to each other one can create the 1D ring.
Magnetic flux applied through the opening of the ring gives rise
to the equilibrium persistent current. In our paper \cite{Nemeth}
we have calculated the persistent current in the 1D ring with a
single $\delta$ barrier. We have applied the same Hartree-Fock
model as in this work, except that the Hartree-Fock equation with
cyclic boundary condition \cite{Nemeth} is the eigenvalue problem
rather than the tunneling problem.

\begin{figure}[tb]
\begin{center}
\includegraphics[clip,width=12cm]{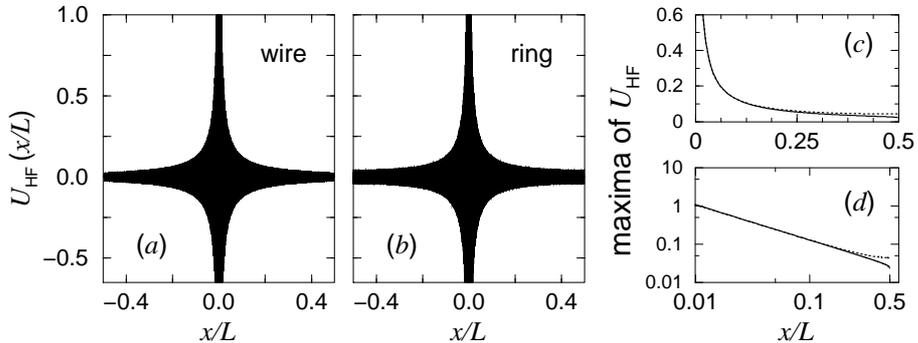}
\end{center}
\vspace{-0.6cm}
\caption{{\small{ (\emph{a}) The $U_{HF}$ data from the Fig~\ref{Fig:2}.
(\emph{b}) Analogous data for the 1D ring from our paper
\cite{Nemeth}. (\emph{c}) Maxima of $U_{HF}$ for the wire (full
line) and ring (dashed line). (\emph{d}) The same as in the panel
(\emph{c}) but in log scale.}}} \label{Fig:3}
\end{figure}

\begin{figure}[tb]
\begin{center}
\includegraphics[clip,width=10cm]{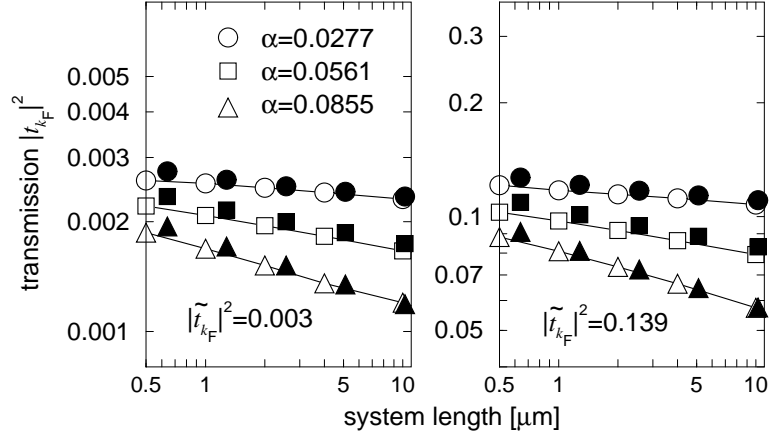}
\end{center}
\vspace{-0.6cm}
\caption{{\small {The transmission probability for the 1D wire (the open
symbols represent the data from Fig.~\ref{Fig:1}) compared with
the transmission probability extracted from the persistent current
of the 1D ring (full symbols). The left and right panels show the
transmission probability for $\left| \tilde{t}_{k_F} \right|^2=0.003$
and $\left| \tilde{t}_{k_F} \right|^2=0.139$, respectively.}}}
\label{Fig:4}
\end{figure}

In Fig.~\ref{Fig:3} we compare the Hartree-Fock potentials in
the wire and ring. Obviously, the amplitude of the Friedel
oscillations in the ring saturates at the boundaries. However, the
amplitude of the Friedel oscillations in the wire decays with
distance from the scatterer without any tendency to saturate (note
in panel (\emph{d}), that the decay of the full line in log scale
is linear or even slightly faster than linear). Nevertheless, in both
cases $U_{HF}(y)$ involves the scaling factor $(d/L)^{\alpha}$ and the
``envelope" scales to a single curve for all $L$. It is thus not surprising,
that also the persistent current in the ring ($I$) comes out from
the Hartree-Fock model~\cite{Nemeth} as a power law, namely $LI
\propto (d/L)^\alpha$. This suggests that the conductance of the
interacting wire might be obtainable from the persistent current
in the interacting ring. In the non-interacting case $I = (ev_F/2
L) |\tilde{t}_{k_F}|$ for magnetic flux $0.25 h/e$. Applying this
formula (intuitively) to the interacting electrons, we obtain the
wire conductance from the formula $|{t}_{k_F}|^2 = 2LI/ev_F$, with
$I$ taken from the persistent current simulation~\cite{Nemeth}. In
Fig.~\ref{Fig:4} we show that the results (full symbols)
reasonably agree with our direct calculation of the wire
conductance.

In conclusion, using the self-consistent Hartree-Fock
approximation at zero temperature, we have calculated the Landauer
conductance of the weakly-interacting spinless electron gas in a
1D wire with a single $\delta$ barrier. We have found the
universal power law $\left| t_{k_F} \right|^2 \propto
(d/L)^{2\alpha}$ known from the Luttinger-liquid model
\cite{Kane-92} and RG models \cite{Matveev-93,Meden-02}. We have
also found that essentially the same wire conductance can be
extracted from the persistent current in a 1D ring.

We thank for the APVT grant APVT-51-021602 and VEGA grant
2/3118/23.


\end{document}